\documentclass[aps,prd,groupedaddress]{revtex4}
\usepackage{graphicx}

\input epsf

\def\p0{\phi_{0}}

\def\hatn{{\bf \hat n}}

\def\vecx{{\bf x}}

\def\veck{{\bf k}}

\def\cP{{\cal P}}
\long\def\comment#1{}

\def\eg{{\it e.g.~}}

\def\W2{{\cal W}}

\def\vecx{{\mbox{\boldmath $x$}}}

\newcommand{\wjma}[6]{\left(
                           \begin{array}{ccc}
         #1 & #2  & #3  \\
         #4 & #5  & #6
                           \end{array}
                   \right)}
\newcommand{\bi}{B_{l_1 l_2 l_3}}

\def\ben{\begin{enumerate}}
\def\een{\end{enumerate}}
\def\bi{\begin{itemize}}
\def\ei{\end{itemize}}
\def\be{\begin{equation}}
\def\ee{\end{equation}}
\def\bea{\begin{eqnarray}}
\def\eea{\end{eqnarray}}

\def\simless{\mathrel{\mathpalette\fun <}}

\def\cmm2{{\,\rm cm^{-2}}}
\def\cm2{{\,{\rm cm}^2}}
\def\cmm3{{\,{\rm cm}^{-3}}}
\def\gcmm3{{\,{\rm g\,cm^{-3}}}}

\def\fun#1#2{\lower3.6pt\vbox{\baselineskip0pt\lineskip.9pt
  \ialign{$\mathsurround=0pt#1\hfil##\hfil$\crcr#2\crcr\sim\crcr}}}

\begin{document}

\date{\today}
\title{The Signature of Patchy Reionization in the Polarization Anisotropy of the CMB}

\author{Olivier Dor\'e$^{(1)}$, Gil Holder$^{(2)}$, Marcelo
  Alvarez$^{(3)}$, Ilian T. Iliev$^{(1)}$, Garrelt Mellema$^{(4)}$,
  Ue-Li Pen$^{(1)}$, Paul R. Shapiro$^{(5)}$}
\affiliation{(1) Canadian Institute for Theoretical Astrophysics, University
  of Toronto, 60 St. George Street, Toronto, ON M5S 3H8, Canada\\
(2) Department of Physics, McGill University, Montreal, QC H3A 2T8, Canada\\
(3) Kavli Institute for Particle Astrophysics and Cosmology, Stanford
  University, P.O. Box 20450, MS 29, Stanford, CA, 94309, USA\\
(4) Stockholm Observatory, AlbaNova University Center, Stockholm
  University, SE-106 91 Stockholm, Sweden\\
(5) Department of Astronomy, University of Texas, Austin, 78712-1083, USA}

\begin{abstract}
The inhomogeneous ionization state of the universe when the first
sources of ionizing radiation appeared should lead to anisotropies
in the polarization of the cosmic microwave background. We use
cosmological simulations of the process by which the first sources
ionized the intergalactic medium to study the induced polarization
anisotropies. We find that the polarization anisotropies have rms of order 
$\sim0.01\,\mu K$, and local peak values of $\sim0.1\,\mu K$,
smaller than those due to gravitational lensing on small
scales. The polarization direction is highly coherent over degree 
scales. This directional coherence is not expected from either primary
anisotropy or gravitational lensing effects, making the largest signals
due to inhomogeneous ionization relatively easy to isolate, should
experiments achieve the necessary very low noise levels.
\end{abstract}

\maketitle

\renewcommand{\thefootnote}{\arabic{footnote}}
\setcounter{footnote}{0}

\section{Introduction}

The cosmic microwave background (CMB) has greatly enhanced our understanding of 
cosmology in the last decade through precise measurements of the
conditions of the universe at $z \sim 1100$. 
As new CMB polarization experiments target
higher resolution and better sensitivity astrophysical processes at
intermediate redshifts ($z \simless 30$) will become increasingly important
as they leave their faint imprints and distortions in the CMB.

This task seems well underway on the large scales, where linear
perturbation is valid. The unambiguous signature of reionization on those scales
has been well measured as an excess of powers in the so-called E modes generated
by scalar and tensor perturbations to the background metric (\citet{page06}). But
experiments currently being built or designed aim at detecting the
fainter B modes, generated only by tensor perturbation to the
metric. However, at smaller angular scales, the picture is more complicated and less certain. All types of perturbations
beyond linear order produce both E and B modes and gravitational
lensing, for example, will transfer power from E to B
modes. Furthermore, at smaller scales, the details of the reionization
history will matters critically. As such, controling exactly the extent to which those
non-linear term contribute to the CMB polarization signal is an
important task we adress in this paper, focusing on the signature of
reionization at small angular scales.

One possible source of polarization anisotropy at small scales is
indeed Thomson scattering by free electrons at the time when the universe goes 
through a transition from neutral to ionized at some time $z>6$. 
The inhomogeneous distribution of ionized and neutral patches might be
expected to provide an enhanced level of inhomogeneity in the Thomson
optical depth; through Thomson scattering of the quadrupole
component of the anisotropy of the CMB it could be possible to get
significant CMB polarization anisotropy on small scales. 
Estimates of this signal using semi-analytic techniques 
(\citet{weller99, hu00, liu01}) suggested that this signal is 
likely to be subdominant. Recent estimates \citet{mortonson06} largely 
agree with this prognosis, but there is certain disagreement between 
simulations and analytic models of other reionization observables like
redshifted 21-cm line of hydrogen and CMB temperature anisotropies 
from kinetic Sunyaev-Zel'dovich effect 
(\citet{21cmreionpaper,2006NewAR..50..909I,kSZ}). For example, recent 
simulations (\citet{2006NewAR..50..909I,kSZ}) found a surprisingly high 
contribution to the expected level of momentum transfer to CMB photons 
due to bulk motions in the partially polarized epoch of reionization, 
suggesting that the bubble size distributions assumed in the analytical 
reionization models are unrealistic and that perhaps scattering effects 
in general may have been underestimated.

A large source of complication for understanding CMB polarization 
measurements will be gravitational lensing of the primary CMB polarization
pattern. To first order,
gravitational lensing does not rotate the plane of polarization
but does distort the CMB anisotropy spatial pattern and induce higher
order correlations in an intrinsically Gaussian distribution. These higher
order correlations can be used to estimate the level of distortion caused
by gravitational lensing and thus remove the induced distortions that
could otherwise mimic the imprint of gravitational radiation generated
during an initial period of cosmic inflation (\citet{knox02, seljak05}). 
Patchy reionization could provide a significant source of noise for this cleaning process and could
lead to a fundamental floor to the accuracy to which we could measure
gravitational radiation signatures in the CMB.

In this work we use recent simulations of reionization that
include self-consistent radiative transfer during the entire reionization
process coupled with a accurate treatment of the coherence of the
primordial quadrupole anisotropy to calculate the expected 
polarization  anisotropy of the CMB due to the reionization of the
universe.

In the next section we describe our simulations, followed by
a description of the numerical techniques for generating polarization
maps from the simulation outputs. We then discuss the levels of
polarization that we find and compare with analytic estimates. We follow
this with a discussion of implications for cleaning of gravitational
lensing effects and close with a summary and discussion of future
directions.

\section{Simulations}
\label{simul_sect}

Our simulations follow the evolution of a comoving simulation volume
of $(100\,h^{-1}\rm Mpc)^3$. Our basic methodology and simulation 
parameters were described in detail in \citet{2006MNRAS.369.1625I},
\citet{mellema06} and \citet{selfregulated}. Here we provide just 
a brief summary. First we perform a very large pure dark matter simulations 
of early structure formation, with $1624^3\approx4.3$ billion particles and 
$3248^3$ grid cells \footnote{$3248=N_{nodes}\times(512-2\times24)$, where 
$N_{nodes}=7$ (with 4 processors each), 512 cells is the Fourier transform 
size and 24 cells is the buffer zone needed for correct force matching on 
each side of the cube.} using the particle-mesh code PMFAST 
(\citet{2005NewA...10..393M}). Such a high mass resolution allows for reliable 
identification (with 100 particles or more per halo) of all collapsed dark matter 
halos with masses $\sim2\times10^9M_\odot$ or larger. We save the detailed 
halo catalogs, which contain the halo positions, masses and properties, in 
up to 100 time slices starting from high redshift ($z\sim30$) down to the 
observed end of reionization at $z\sim6$ or later. We also save the 
corresponding density and bulk peculiar velocity fields at the resolution of the 
radiative transfer grid. Since radiative transfer simulations at the full grid 
size of our N-body simulations are not practical, we follow the radiative 
transfer on coarser grids, of sizes $203^3=(3248/16)^3$ or $406^3$. 

For this study we assume two flat $\Lambda$CDM cosmologies, the first 
with parameters 
($\Omega_m,\Omega_\Lambda,\Omega_b,h,\sigma_8,n)=(0.27,0.73,0.044,0.7,0.9,1)$
\citep[][hereafter WMAP1]{2003ApJS..148..175S}, and 
($\Omega_m,\Omega_\Lambda,\Omega_b,h,\sigma_8,n)=(0.24,0.76,0.042,0.73,0.74,0.95)$ 
\citep[][hereafter WMAP3]{2006astro.ph..3449S}, where $\Omega_m$, 
$\Omega_\Lambda$, and $\Omega_b$ are the total matter, vacuum, and baryonic 
densities in units of the critical density, $h$ is the Hubble constant in 
units of 100~$\rm km\,s^{-1}Mpc^{-1}$, $\sigma_8$ is the standard deviation 
of linear density fluctuations at present on the scale of 
$8 \rm h^{-1}{\rm Mpc}$, and $n$ is the index of the primordial power 
spectrum. We use the CMBfast transfer function (\citet{1996ApJ...469..437S}). 

All identified halos are assumed to be sources of ionizing radiation and 
each is assigned a photon emissivity proportional to its total mass, $M$, 
according to
\be 
  \dot{N}_\gamma=f_\gamma\frac{M\Omega_b}{\mu m_p t_s\Omega_0},
\ee
where $t_s$ is the source lifetime, $m_p$ is the proton mass, $\mu$ is the 
mean molecular weight and $f_\gamma$ is an assumed photon production efficiency 
which depends on the number of photons produced per stellar atom, the star 
formation efficiency (i.e. what fraction of the baryons are converted into stars) 
and the escape fraction (i.e. how many of the produced ionizing photons escape the 
halos and are available to ionize the IGM).

The radiative transfer is followed using our fast and accurate ray-tracing
photoionization and non-equilibrium chemistry code C$^2$-Ray (\citet{methodpaper}). 
The code has been tested in detail for correctness and accuracy against available 
analytical solutions and a number of other cosmological radiative transfer codes
(\citet{methodpaper,2006MNRAS.371.1057I}). The radiation is traced from every 
source on the grid to every cell using short-characteristic ray-tracing.
 
We have performed four radiative transfer simulations with WMAP1 background 
cosmology and two simulations with WMAP3 parameters. Each of these sets share 
the source lists and density fields given by the underlying N-body simulation, 
but adopt different assumptions about the source efficiencies and the sub-grid 
density fluctuations. The WMAP1 and WMAP3 simulations use different random seeds. 
The runs and notation are the same as in \citet{mellema06} and 
\citet{selfregulated}. Simulations f2000 and f250 assume $f_\gamma=2000$ and 250, 
respectively, and no sub-grid gas clumping, while f2000C and f250C adopt the same 
respective efficiencies, $f_\gamma=2000$ and 250, but also add a sub-grid gas 
clumping, $C(z)=\langle n^2\rangle/\langle n\rangle^2$, which evolves with redshift 
according to
\be
C_{\rm subgrid}(z)=27.466 e^{-0.114z+0.001328\,z^2}.
\label{clumpfact_fit}
\ee
in WMAP1 cosmology and as
\be
C_{\rm sub-grid}(z)= 26.2917e^{-0.1822z+0.003505\,z^2}.
\label{clumpfact_fit3}
\ee
for WMAP3 cosmology. These fits were obtained from another two 
high-resolution PMFAST N-body simulation, with box sizes 
$(3.5\,\rm h^{-1}~Mpc)^3$ and a computational mesh and number of particles 
of $3248^3$ and $1624^3$, respectively. These parameters correspond to a 
particle mass of $10^3M_\odot$ and minimum resolved halo mass of 
$10^5M_\odot$. This box size was chosen so as to resolve the scales most 
relevant to the gas clumping - on scales smaller than these the gas 
fluctuations would be below the Jeans scale, while on larger scales the 
density fluctuations are already present in our computational density 
fields and should not be doubly-counted. The expression in 
equation~(\ref{clumpfact_fit}) excludes the matter inside collapsed 
minihalos (halos which are too small to cool atomically, and thus have
inefficient star formation) since these are shielded, unlike the generally 
optically-thin IGM. This self-shielding results in a lower contribution of 
the minihalos to the total number of recombinations than one would infer 
from a simple gas clumping argument
(\citep{2004MNRAS.348..753S,2005MNRAS...361..405I,2005ApJ...624..491I}). The 
effect of minihalos could be included as sub-grid physics as well, see 
(\citet{MH_sim,2006astro.ph.10094M}). This results in slower propagation of the 
ionization fronts and further delay of the final overlap. The halos that can 
cool atomically are assumed here to be ionizing sources and their recombinations 
are thus implicitly included in the photon production efficiency $f_\gamma$ 
through the corresponding escape fraction. Such treatment would tend to 
overestimate the effects from small-scale clumping (except for the minihalo 
effects) since we assume that the gas follows the dark matter distribution 
at small scales and we ignore the gradual smoothing of the gas density field 
due to pressure effects in the ionized regions. Thus, the cases with and without 
sub-grid clumping should be considering as bracketing the effect of gas clumping
from above and below.

\section{Propagating linearly polarized CMB photon planes in the simulation}

Linear polarization of the CMB is generated by the Thomson
scattering of incoming radiation with a quadrupole anisotropy
radiation pattern (see for example the general
formalism for CMB polarization described by \citet{zaldarriaga97}). 
In principle we can distinguish three different origins for the
quadrupole anisotropies: primordial from projection of the Sachs-Wolfe
temperature, intrinsic quadrupole from scattering (projection of the
Doppler effect) and kinematic quadrupole from second-order Doppler
effects. We will, however, 
neglect the latter two following the results and arguments from \citet{hu00}.

As we will show below, the quadrupole anisotropy in the radiation
is spatially slowly varying, so the dominant source of small scale fluctuations
in the polarization are from fluctuations in the number of electrons that are the
targets for Thomson scattering.
We use the simulations to provide the free electron distribution;
to calculate the polarization anisotropy we must calculate the  
quadrupole for each $z$ and its correlation properties over large scales. Past work
investigating polarization anisotropies due to late time Thomson scattering
has often assumed a constant quadrupole; a notable exception is the work of 
\citet{amblard05}, where they used a very low resolution realization of the universe
to keep track of CMB anisotropy at different locations. We instead use analytic
tools to keep track of the covariance between observed quadrupoles at different
locations and epochs in the universe.  

\subsection{Quadrupole as a function of time and Q,U maps generation}

We closely follow \citet{hu00} and
\citet{portsmouth04} in our formalism. 
The polarization generated at a position $\vecx$
($\hat\vecx$ the associated unit vector), with the observer being at
the origin of the reference frame, and $\tau$ being the conformal
lookback time ($\tau=0$ today and $\|\vecx\|=\tau$) can be written as
\bea
\left[Q\pm iU\right](\hatn) & = & {1\over 10}\sqrt{3\over4\pi}\int_0^\tau d\chi\ g\sum_{m=-2}^2 a_{2m}(\vecx,\chi)  _{\pm 2}Y_2^m(\hatn)\\
g(\vecx,\chi) & = & \dot\tau_T(\chi) e^{-\tau_T(\chi)}
\label{eq:qu_map}
\eea
where ${}^{}_{\pm 2}Y_{2m}^{}$ are defined as \eg in \citet{hu97}. The
optical depth to Thomson scattering $\tau_T$ is given by  
\bea
\tau_T(\tau) = \sigma_T \int_0^\tau n_e(\chi) d\chi
\eea
and the quadrupole at any point is
\bea
a_{2m}(\vecx)       & = & -4\pi\int d^3\veck\ e^{i \veck \cdot
  \vecx}_{}\Delta_2^{}(\veck,\tau)Y_{2m}^*(\hat\veck)\label{eq:a2m}\\
\Delta_\ell^{}(\veck,\tau) & = & \Delta_\ell^{}(k,\tau)\Phi_i^{}(\veck)\\
\Delta_\ell^{}(k,\tau) & = & {3\over 10}j_\ell^{}[k(\tau-\tau_{rec})]+{9\over 5}\int_{0}^{\tau-\tau_{rec}}d\chi'j_\ell^{}(k\chi'){\partial\over
  \partial\tau}\left({D\over a}\right)(\tau-\chi')
\eea
where $\Phi_i$ is the initial gravitational potential and $\Delta_\ell^{}(k,\tau)$ is the transfer function.
Computing $a_{2m}$ at all cells of the simulation volume is computationally prohibitive but fortunately not necessary since the
quadrupole is hardly varying within a 100 h/Mpc box as we show below by computing the auto-correlation of $a_{2m}(\vecx)$. It is
varying from box to box as new modes enter the horizon but
only slowly, since the kernel convolving the gravitational
potential $\Phi_i^{}(\veck)$ in Eq.~\ref{eq:a2m} is broad.

\begin{figure}[t!]
\begin{center}
\includegraphics[width=0.48\textwidth]{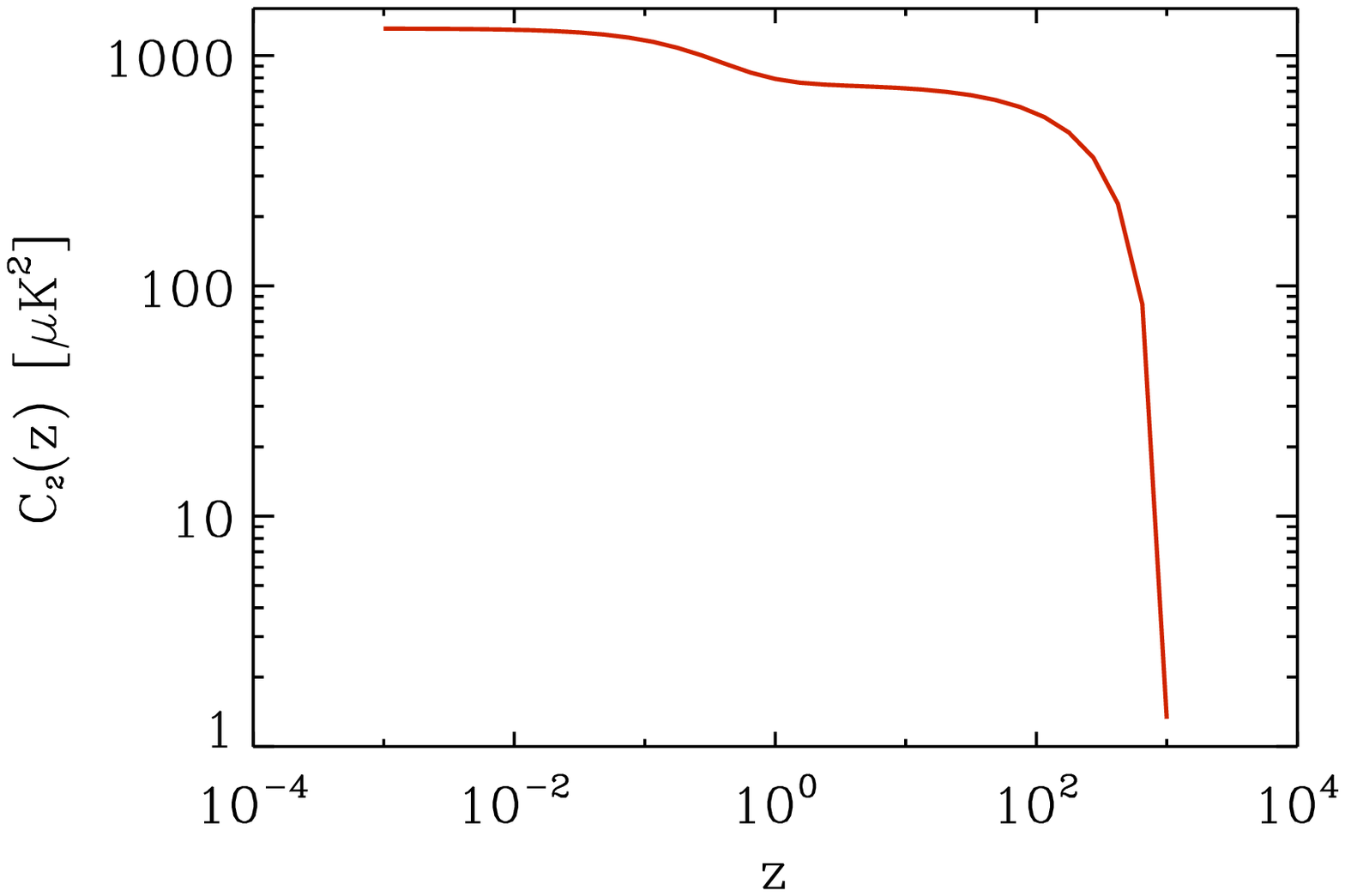}
\includegraphics[width=0.48\textwidth]{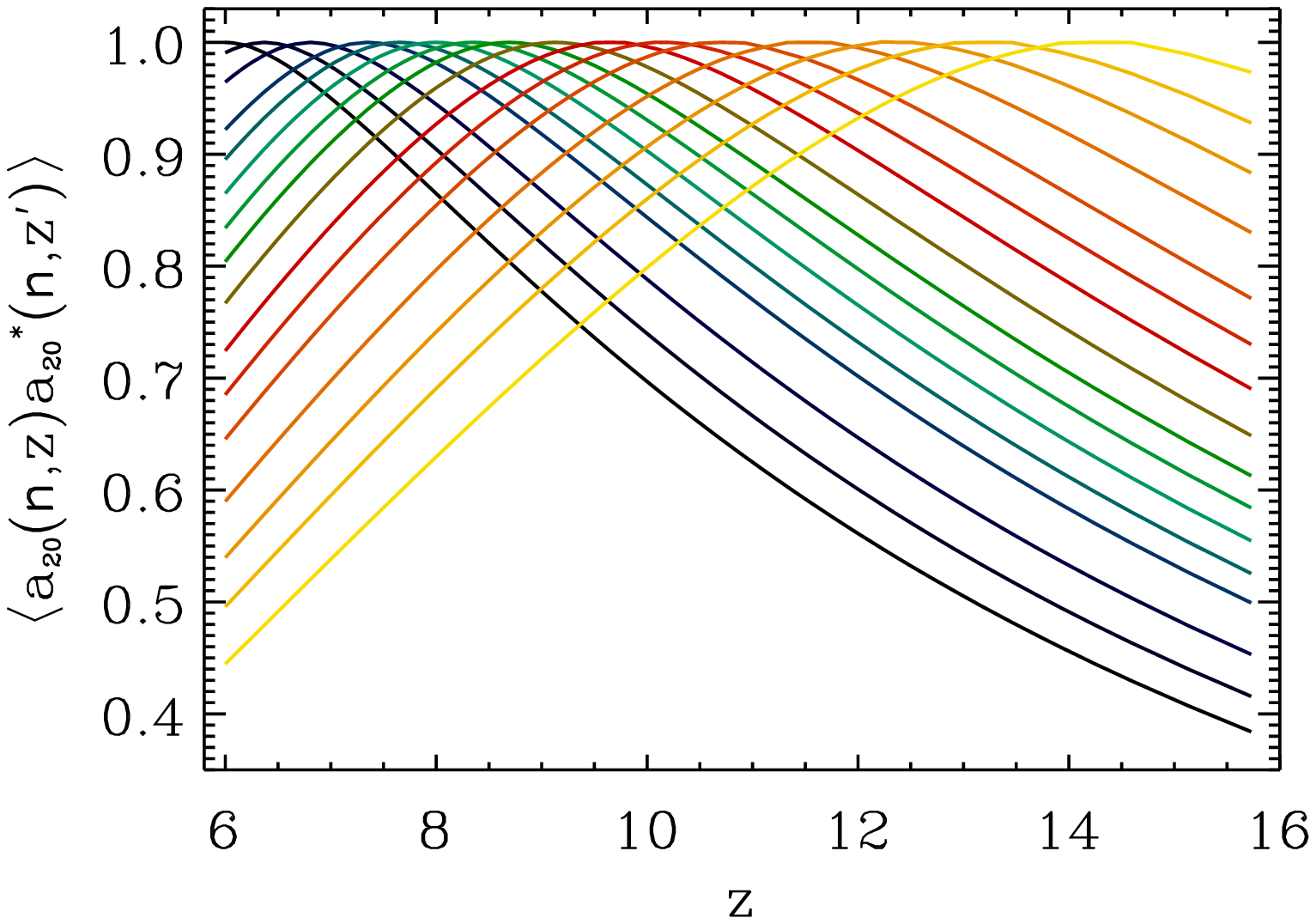}
\end{center}
\caption{\emph{Left panel}: Quadrupole $C_2$ as a function of $z$. \emph{Right
  panel}: Correlation $\langle
  a_{2m}^{}(0,0,\tau(z))a_{2m'}^{*}(0,0,\tau(z'))\rangle$ for  $(m,m')=(2,2)$ where $z$
  and $z'$ correspond to the center of each box. Each line/color
  corresponds to a given $z'$. The correlation peaks to 1 of course when
  $z=z'$. Similar results are obtained for $(m,m')=(1,1)$ or $(0,0)$.}
\label{quad_correl}
\end{figure}

To illustrate this let us compute the covariances $\langle a_{2m}^{}(\vecx)a_{2m'}^{}(\vecx') \rangle$ where $\langle \rangle$
denotes ensemble average. Following \citet{portsmouth04} we obtain for $\vecx \ne \vecx'$
\bea
\lefteqn{\langle a_{2m}^{}(\vecx)a_{2m'}^{*}(\vecx') \rangle =}                              &&\\
& & 5(4\pi)^{5/2}_{}(-1)^m_{}\sum_{\ell=0}^2(-1)^\ell_{}\wjma{2}{2}{2\ell}{0}{0}{0}\wjma{2}{2}{2\ell}{-m}{m'}{m'-m}
K_{2\ell}(\vecx-\vecx',\tau,\tau') Y_{2\ell,m'-m}^*(\hat\vecx-\hat\vecx')\nonumber
\label{eq:cov_a2m}
\eea
with 
\be
K_{\ell}(\vecx,\tau,\tau') = \int k^2dk \Delta_2^{}(k,\tau)\Delta_2^{}(k,\tau')\cP_{\Phi\Phi}^0(k)j_\ell^{}(k|\vecx|)
\ee
and where
\be
\langle\Phi_i^{}(\veck)\Phi_i^{*}(\veck') \rangle = \delta_D(\veck-\veck')\cP_{\Phi\Phi}^0(k)\ .
\ee
whereas we obtain for $\vecx=\vecx'$,
\be
\langle a_{2m}^{}(\vecx)a_{2m'}^{*}(\vecx') \rangle = \delta_{mm'} C_2(\tau)
\ee
with
\be
C_2(\tau) = (4\pi)^2\int k^2dk \Delta_2^{2}(k,\tau)\cP_{\Phi\Phi}^0(k)\ .
\ee

The important results of this section are summarized in Fig.\ref{quad_correl} where we see in
the left panel the evolution of the quadrupole $C_2$ as a function of
redshift, and in the right panel the correlation between $a_{22}(z)$
and $a_{22}(z')$ as seen by observers at various redshifts along our line
of sight. The redshifts were chosen to match the center of the
simulation boxes. We see from this plot that two consecutive
boxes would have a $a_{22}$'s correlated at more than 99\%. However
this correlation drops to 50\% for the most distant boxes. 
This suggests that the polarized signal from patches of ionization at the beginning
of reionization could be significantly misaligned with that generated at the end of
reionization, provided reionization happens relatively slowly. Conversely, if the 
universe makes a transition from mainly neutral to mainly ionized over a small
redshift range then the polarization anisotropy will be sourced by a quadrupole
that is effectively constant in both direction and amplitude.

Since
similar results are obtained for  $a_{20}$'s and  $a_{21}$, we
conclude that we can reasonably consider the $a_{2m'}$s to be constant
within a box. For the sake of accuracy we will however allow them to vary
between boxes but enforce the correct correlation properties. We
do so by considering the $a_{2m}(z)$ as a multivariate Gaussian field
with the previously computed correlation matrix given in Eq.~\ref{eq:cov_a2m}.

Once we generated this set of $a_{2m}(z)$, it is straightforward
applying Eq.~\ref{eq:qu_map} to propagate a photon plane through the
boxes to generate a set of $\tau_T$, $Q$ and $U$ maps. Note that care is taken
during this stage to rotate and translate the boxes by random amounts
to avoid any artifacts due to the fact that the boxes stems from
a single initial condition realization. From these $Q$ and $U$ maps,
we can easily measure the $E$ and $B$ mode power spectra in the flat
sky approximation as well as generate a set of $E$ and $B$ mode maps
(see for example \citet{okamoto02}). 

\begin{figure}
\begin{center}
\includegraphics[width=0.75\textwidth,angle=90]{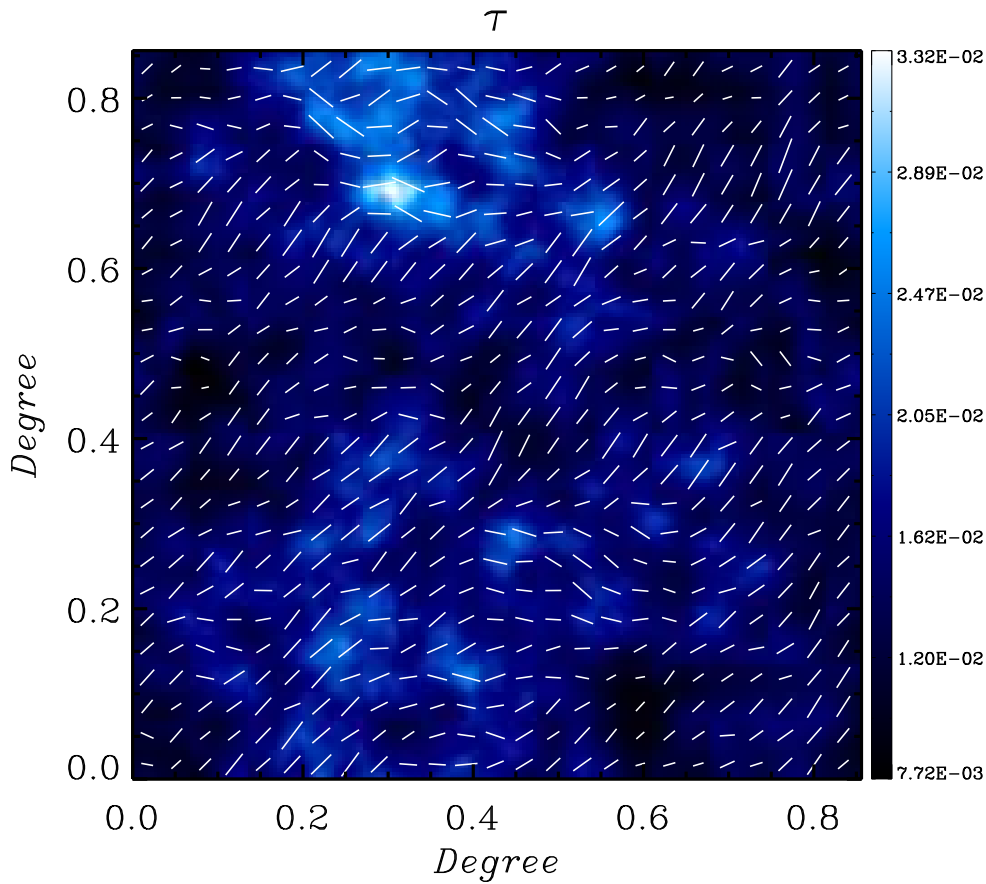}
\end{center}
\caption{Map of optical depth (color scale) and direction of polarization
(shown by sticks denoting the orientation of the electric field) 
at each location in a typical map for the WMAP3 cosmology
and the f250 ionization model. Note that the polarization amplitude is
proportional to the optical depth (and also to the primordial quadrupole
at that location), and that the amplitude of the polarization is {\em not}
proportional to the length of the sticks.
}
\label{fig:taupol}
\end{figure}

\section{Results}

A typical result is shown in Figure~\ref{fig:taupol}.
The optical depth fluctuations are substantial, with peak to peak
variations larger than $\pm 0.01$. Note that these optical depth
fluctuations are comparable to or larger than the Thomson optical
depths along lines of sight through the centers of large galaxy clusters,
the most massive collapsed objects in the local universe. This is 
easy to understand: the central regions of galaxy clusters today are overdense relative to
the background density by factors of order $10^3$ or $10^4$. The universe
at redshift 9 was denser by a factor of $10^3$, so a path length on the
order of 1 physical Mpc will have an optical depth comparable to a 
massive cluster in the local universe. The typical scale of
the fluctuations is arcminutes. The direction of the polarization 
depends on the redshift that is contributing the scattering,
since the quadrupole is highly coherent within a single simulation
box length. Scattering events that are widely separated in redshift should have
polarization angles that are largely uncorrelated. It is evident that
the largest optical depth fluctuations correspond to scattering at
comparable redshifts, although there is a noticeable dispersion in the
polarization angles.

\begin{figure}
\begin{center}
\includegraphics[width=0.9\textwidth,angle=90]{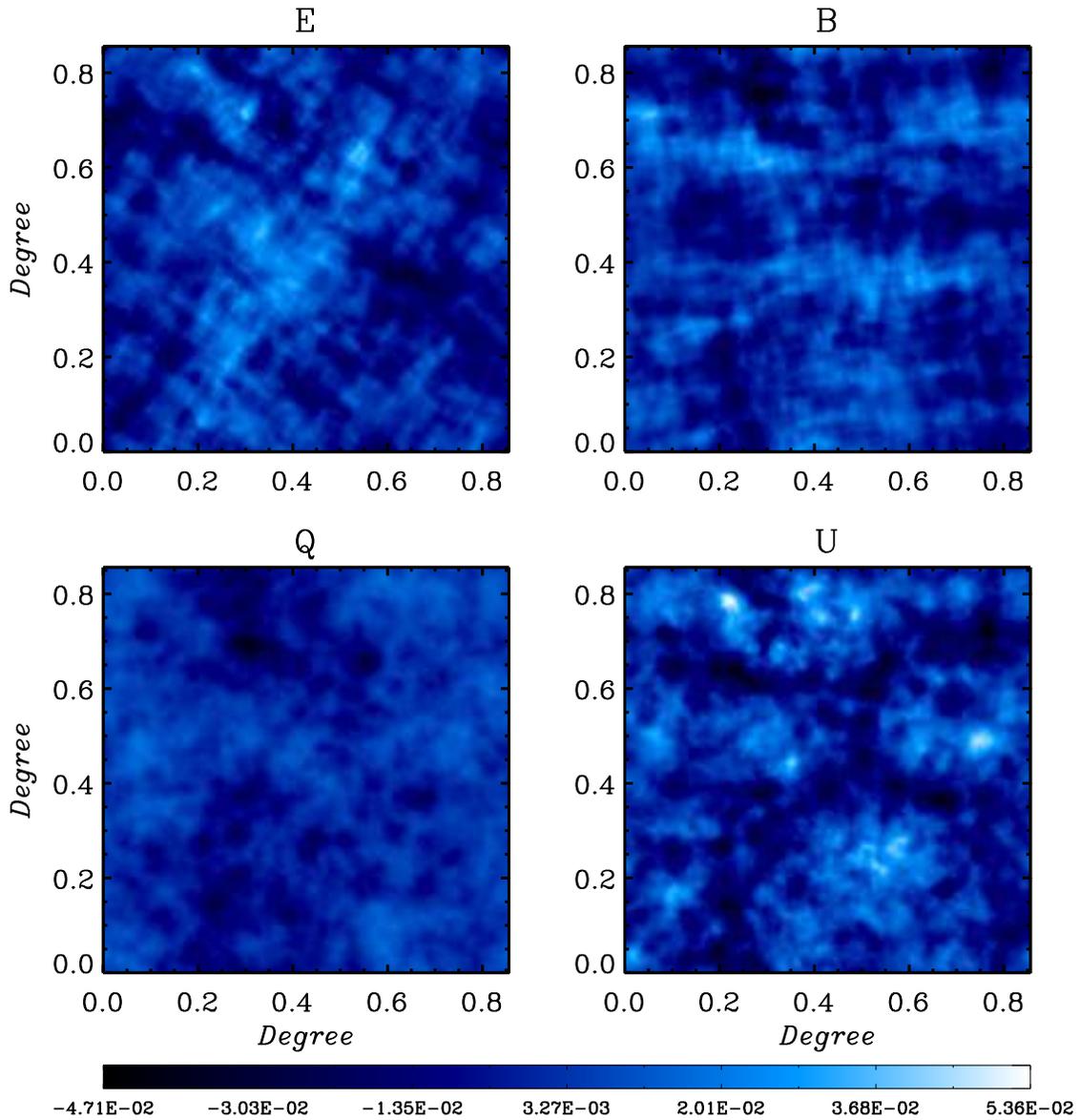}
\end{center}
\caption{Maps showing polarization maps for a typical realization for
the WMAP3 cosmology and the f250 ionization model. Top panels show
E and B maps, while bottom panels show Q and U maps. The mean was
subtracted from each of this map to ease the comparison.
}
\label{fig:ebqu_pol}
\end{figure}

An important question is how this secondary scattering shows up as
E-modes or B-modes. This is shown in Figure~\ref{fig:ebqu_pol}. There is
a difference in the mean values of the two maps, since the first order
scattering effect gives only EE correlations, but the fluctuations around
the mean are of equal amplitudes in the EE and BB maps. By comparison
the bottom panels show the Q and U maps, where it is evident that there
is a large asymmetry, with much more power appearing in one (U in this
case). This is easy to understand: if the scattering was a pure Gaussian
isotropic random field with a coherent quadrupole direction the result
would be exactly equal power in EE and BB, but it would be possible to
choose coordinates such that {\em all} of the power is in a single
Stokes component Q or U (all of the signal with a single polarization
angle). Working in EE and BB confuses the issue in this case.
The coherence of the fluctuations in terms of Q and U suggest that it
should indeed be possible (given sufficient signal to noise) to 
reconstruct the primordial quadrupole at specific locations during
the epoch of reionization, given an estimate of the scattering 
optical depth.

The nature of E and B modes and the apparent excess in U over Q can
be seen in the structure of the E and B maps. The E maps have clear
structures oriented at 45 degrees, while the B maps have a tendency
for structure in the horizontal and vertical directions. By 
construction, the E maps must have the structure varying along
or orthogonal to the direction that the polarization is oriented,
while the B maps will evolve at 45 degrees to the polarization
direction. With most of the polarization being in Stokes U (i.e.,
at $\pm$45  degrees, this requires that the E modes will have structure
mainly banded in the diagonal directions with the B modes showing
banding along the axes. The polarization pattern from reionization
is unlikely to show much resemblance to anything from the early universe.

\begin{figure} 
\begin{center}
\includegraphics[width=0.9\textwidth]{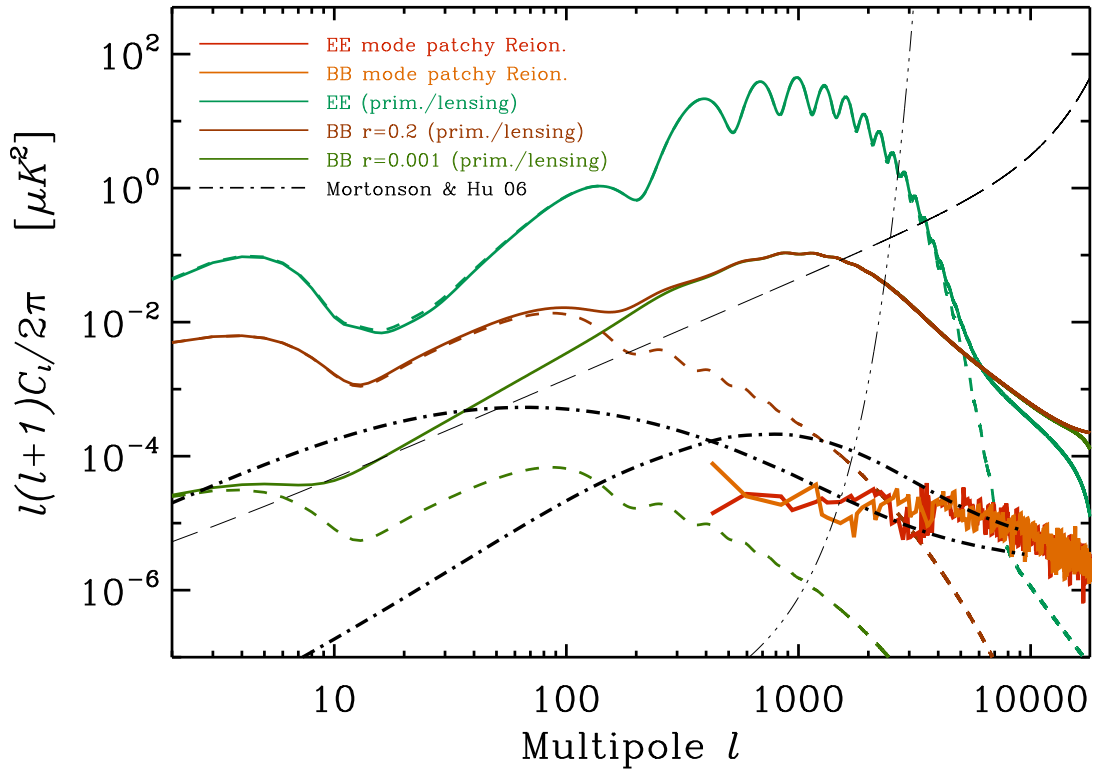}
\end{center}
\caption{Angular power spectrum of EE and BB polarization anisotropy,
compared with primordial EE and BB signals (tensor to scalar ratio, r=0.2) and EE and BB signals including
the effects of gravitational lensing (solid) or not
(dashed). Several analycal computations as in \citet{mortonson06} are shown as
well. The model that peaks around $\ell\simeq 80$ receives most of its
contribution from $\simeq$ 200Mpc (comoving) bubbles whereas the one
that peaks around $\ell\simeq 800$ is dominated by the contribution of
$\simeq$ 40Mpc (comoving) bubbles. Note however than this model is not
meant to reproduce our simulations (see text for details). The power
level is smaller than the lensing-induced BB signal by several  orders
of magnitude on small scales. Planck sensitivity is shown by the
dash-triple dot line, while an estimate for something like ACT or
SPTPol is showm as a dashed line for a bin of width $\Delta
\ell=10$. For comparison, we also display an additional BB spectrum with r=0.001. }
\label{fig:clpol}
\end{figure}

Similarly, the separation in Q and U suggests that this may not be 
a serious impediment for cleaning of lensing. Lensing affects Q and U
equally, so it should be always possible to find at least some modes
that are useful for lensing reconstructions. Some careful
investigations are currently underway to see whether this is indeed
the case. 

This signal might however provide a floor to any ability to measure
either inflationary gravity wave-induced CMB polarization or do
lensing reconstruction. This is shown in Figure~\ref{fig:clpol}. In
this case it appears that these secondary anisotropies will limit
removal of lensing signals beyond the level of 1\% in the power (10\%
in amplitude). The lensing reconstruction depends on the
non-Gaussianity of the signal, so the power spectrum is not necessarily the best
place to do this comparison. The maps are clearly non-Gaussian.

This is in rough agreement with \citet{mortonson06} and
\citet{liu01}, and the level is comparable to that found in \citet{baumann03b}. However, the latter
work looked at the contribution due to scattering in the fully
ionized universe, and is a component that should be added to this
signal. The coherence of the quadrupole in its direction is not
expected for the superposition of scattering extending from the local
universe out to distances corresponding to the epoch of reionization. 

As carefully addressed in \citet{mortonson06} and \citet{liu01}, a
crucial consideration for polarization anisotropy is the size of
the characteristic ionized regions. The larger regions considered by the
former paper are comparable to the simulation volume and thus it is
difficult to explicitly compare results (see
Fig.~\ref{fig:clpol}). Note that those results would certainly limit
our ability to measure at low multipoles primordial B modes with a tensor to scalar
ratio around 0.005. However, such large scale bubbles do not occur in our simulations. On the other hand, if the bubble distribution of
\citet{mortonson06} is made to reproduce the one observed in our simulations --by
typically adjusting the properties of the distribution of bubbles and
its evolution-- then the analycal computations can be in reasonnable
agreement with ours (see Fig.~\ref{fig:clpol}). We leave a more detailed
comparison between this analytical model and our results for future work.

We find somewhat more small
scale power, but the simulations don't have sufficient volume to 
accurately probe degree scale anisotropy. Our calculation is similar to
that of \citet{liu01}, but with a more careful treatment of the process
of reionization and an evolving quadrupole. Note that ignoring the
evolution of the quadrupole will lead to nearly exactly equal EE and
BB polarization, since the generated anisotropy will have the same plane
of polarization and there is a coordinate system in which the signal is
purely Stokes Q or Stokes U. A statistically isotropic amplitude distribution
with a fixed polarization direction will necessarily have an equal number of
modes parallel or perpendicular to the polarization direction as at $\pm$45 
degrees.

\section{Discussion}

\begin{figure}
\begin{center}
\includegraphics[width=0.3\textwidth]{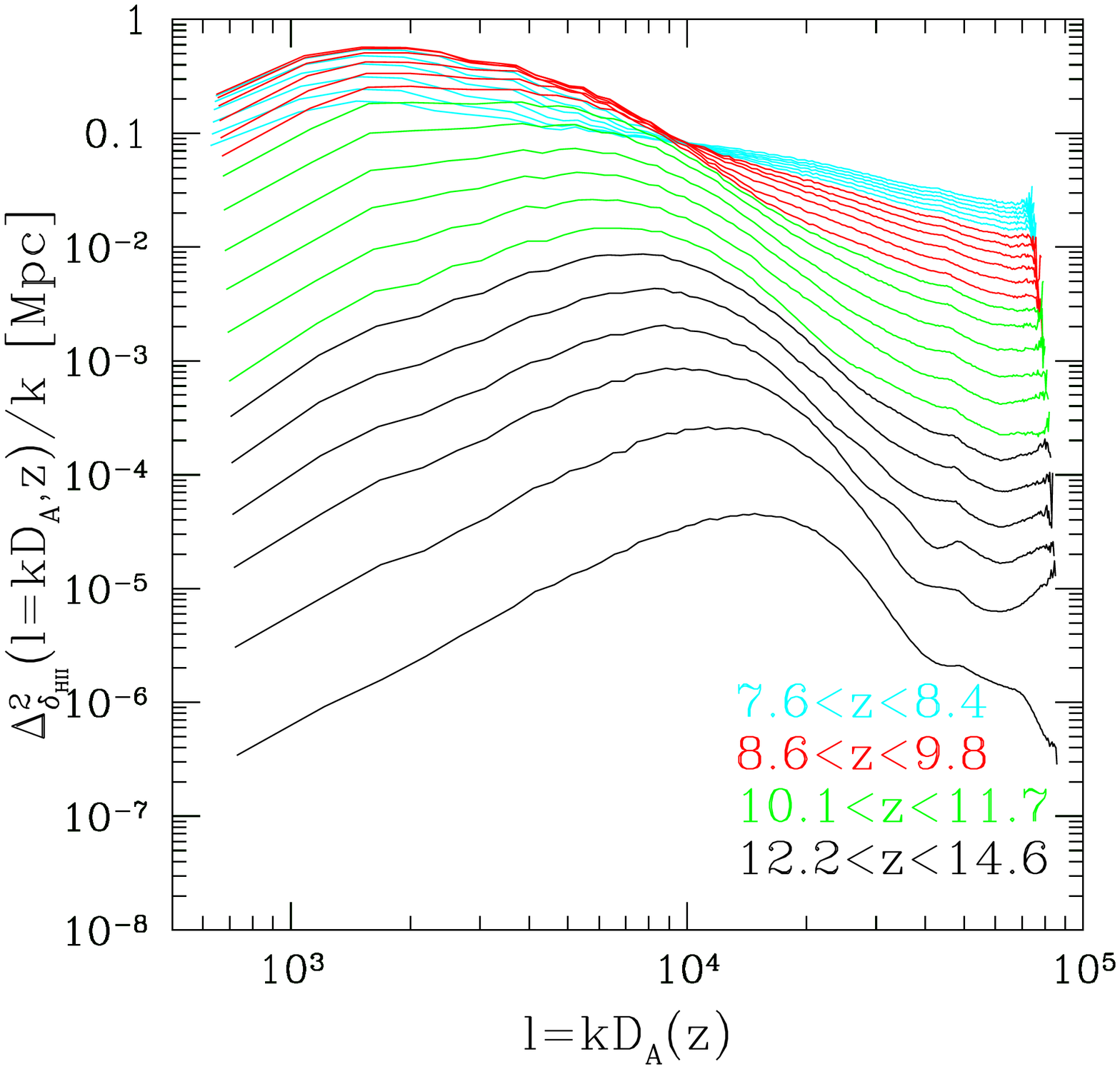}
\includegraphics[width=0.3\textwidth]{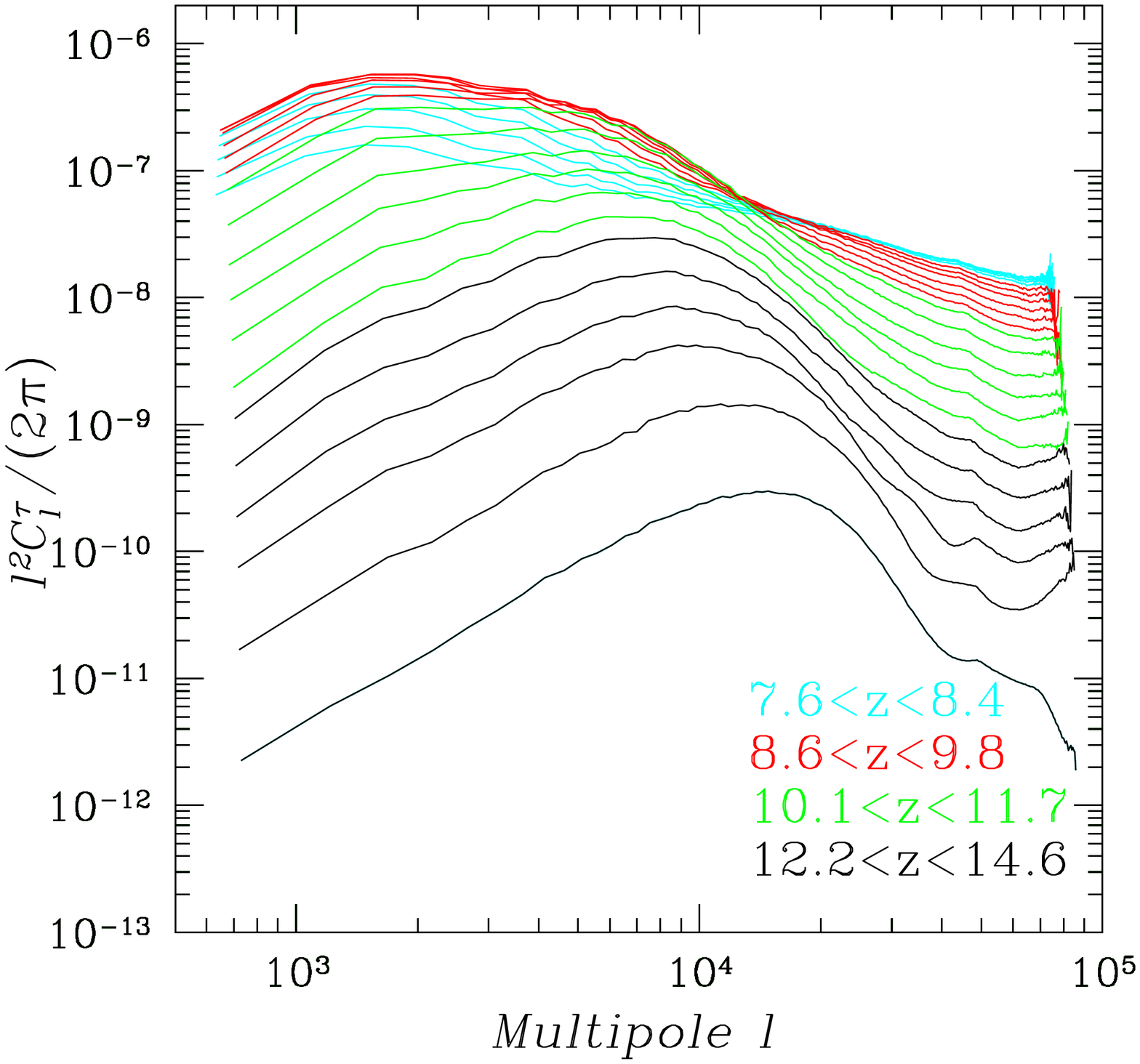}
\includegraphics[width=0.33\textwidth]{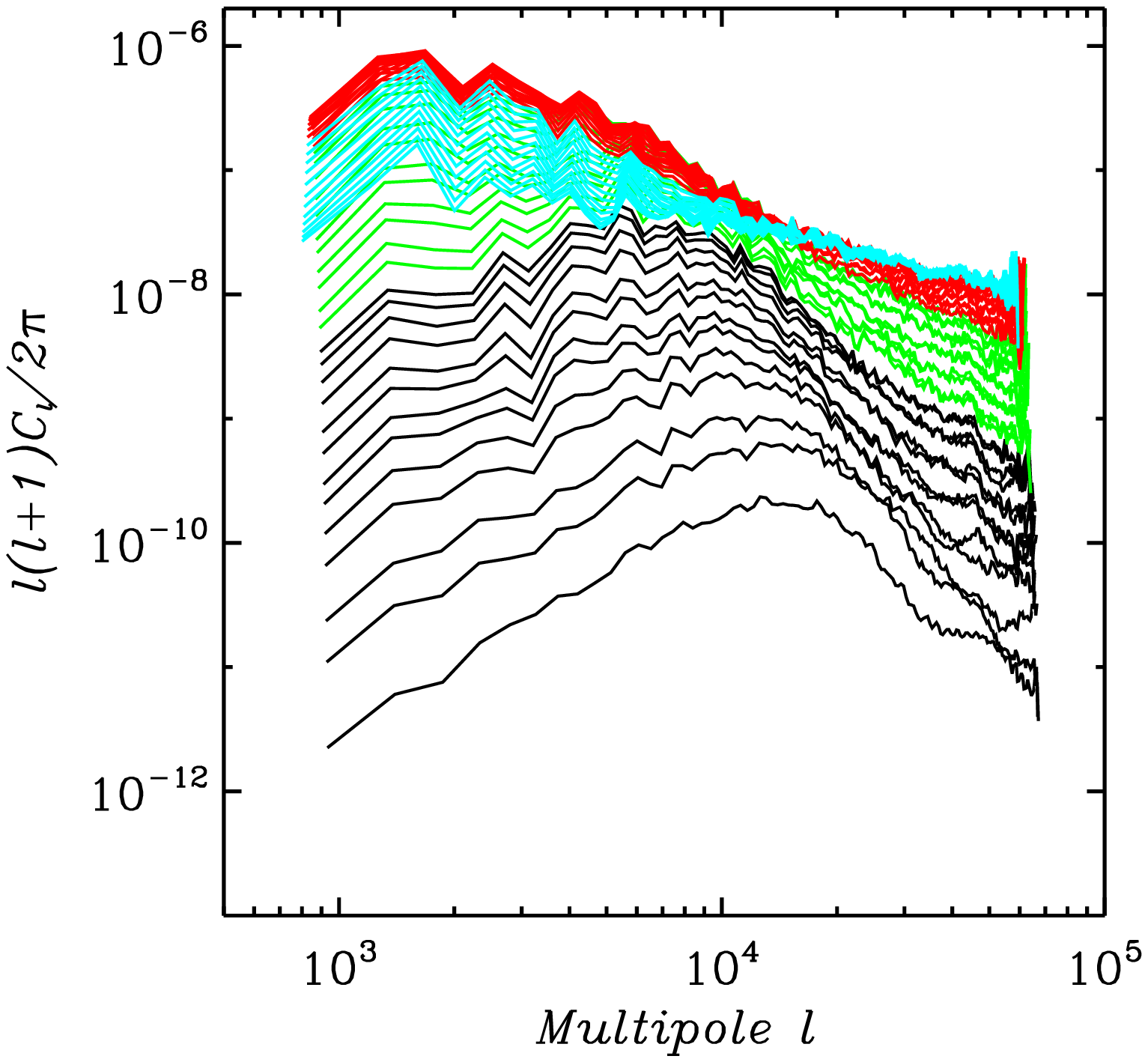}
\end{center}
\caption{Measured angular power spectra of individual simulation volumes at different
redshifts, showing (left panel) the projection of the 3D power spectrum of the
ionized hydrogen distribution using the Limber approximation, (middle) the
angular power spectrum of the optical depth maps at each redshift
deduced from this 3D power spectrum and (right) the measured angular
power spectra for each redshift.}
\label{fig:evol_ne}
\end{figure}

The {\em rms} polarization signal on small scales due to 
the inhomogeneous process of the ionization of the universe
at $z \sim 10$ is relatively small, on the order of 0.01 $\mu K$,
much smaller than the anisotropy induced by gravitational lensing
of the primordial fluctuations. The signal has a unique signature,
in that the polarization angle should be relatively constant over
degree scales, with amplitude variations on scales of several arcminutes.
This is a highly non-generic polarization anisotropy and is not what is
expected from either gravitational lensing or primordial fluctuations.
This is in line with previous expectations, although the non-Gaussianity
of the optical depth distribution leads to some surprisingly large
fluctuations at a few locations (approaching 0.1 $\mu K$). 
This directional coherence should make the largest signals
due to inhomogeneous ionization relatively easy to isolate, should
experiments achieve the necessary very low noise levels. Note that
this separation will also benefit from the mapping of E modes on large
scales (B modes signal is expected to be weaker there)  where the
reionization signature is averaged over the size of the horizon (\citet{yadav05}). The
small scale patchy reionization would appear as an excess of power on
smaller scales but with the same direction.

The signal will provide an ambitious target for upcoming CMB experiments.
Noise curves for Planck
\footnote{\texttt{http://www.rssd.esa.int/Planck}}($\sigma=$2.3$\mu$K/5.0'fwhm,
0.85\% of the sky) and rough estimates for ACT (Atacama Cosmology 
Telescope; \citet{kosowsky06}) and SPTPol (South Pole Telescope;
\citet{ruhl04}) ($\sigma=$2.0$\mu$K/1.7'fwhm, 0.25\% of the sky)
are shown in Figure~\ref{fig:clpol}. The noise levels are orders of magnitude
above what is 
required to image the polarization from inhomogeneous reionization. However,
the non-Gaussianity of the signal means that it may be possible to detect some
of this signal with near future instrumentation. Our computations also
suggest that experiments targetting large scale measurements of the BB
modes (\eg SPIDER or EBEX) should not be concerned about this source of
contamination. Note that our simulation size does not allow nor
support the kind of very large bubbles considered by \citet{mortonson06} as
displayed in Figure~\ref{fig:clpol}.

If we were to consider a dedicated survey, a very wide dish like the Large Millimeter Telescope 50m dish
($\Theta_{fwhm}=0.07'$ at 100GHz) would be appropriate. Since our signal is about
$\sigma=$0.01$\mu$K, and since current millimeter detectors achieve sensitivies of 200$\mu
K\sqrt{s}$ at 100GHz, observing our signal with a signal to noise of
1 per beam would require observing for $N_{days}$ days with
$N_{det}$ detectors with $N_{days}N_{det}\simeq 4600$. Note that
focal planes being currently developped encompass several thousand detectors.

Alternatively, an independent and complementary approach proposed recently for deriving 
the optical depth fluctuations from patchy reionization is by using good 
quality redshifted 21-cm maps when such maps become available in the future
(\citet{pol21}).

Even with large non-Gaussianity, the Limber approximation 
(\citet{kaiser92, limber53}) is still sound. This is evident
from Figure~\ref{fig:evol_ne}. The left panel essentially shows the result
of calculating the 3D power spectrum of each box and using the Limber
approximation to estimate the angular power spectrum while the right panel
first projects the ionized distribution and then calculates the angular power
spectrum directly. As long as the ionization distribution is relatively constant
over the light crossing time of the typical fluctuations then one expects the Limber
approximation to hold. 
We don't allow evolution within a single box as we make
the maps, so it is possible that there is some additional signal coming from
regions where the ionization structure is evolving rapidly. This may be of
particular importance when regions are seeing several ionizing sources
simultaneously.

\begin{figure}
\begin{center}
\includegraphics[width=0.9\textwidth]{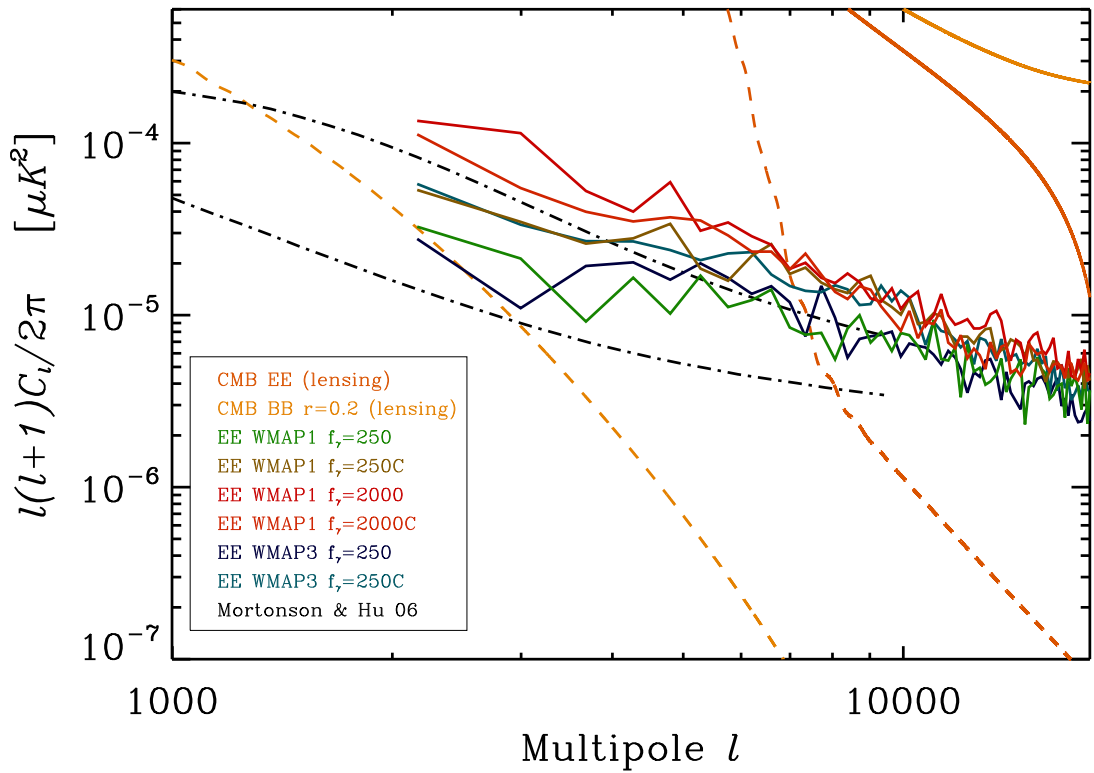}
\end{center}
\caption{Angular power spectra of EE polarization generated with
  various assumptions concerning either the cosmology (WMAP1 versus
  WMAP3 model) or the physics of reionization (various photon production
efficiency, $f_\gamma$, and with or without sub-cell gas clumping), as labelled 
by color. The primordial CMB is in solid line whereas the lensed one
  is in solid line. Analytics computation as in Fig.~\ref{fig:clpol}.
}
\label{fig:cl_comparison}
\end{figure}

As we discussed in detail in Section~\ref{simul_sect}, we simulate several
reionization scenarii with varying photon production efficiency, $f_\gamma$,
level of sub-cell gas clumping and background cosmology. The effect of these
variations on the polarization signal is displayed in
Fig.\ref{fig:cl_comparison}.  The results prove fairly robust and largely
dependent of the background cosmology.  However, some interesting dependence
on the reionization scenarios is observed.  In particular, the polarization
signal is higher (by factor of $\sim2-3$ in $C_\ell$'s, for most of the range,
but up to an order of magnitude at $\ell\sim4000$) in extended-reionization
scenarios (f250C cases) or high source efficiency (corresponding to large
ionized bubble sizes) scenarios (f2000 and f2000C) compared to scenarios which
lack both of these characteristics (f250 cases). Furthermore, the high source
efficiency scenarios have somewhat more power at large scales, as expected.
All these variations are relatively small, however, which combined with the
weakness of the polarization signal makes it unlikely that polarization
measurements would be used to distinguish different reionization scenarios.

The results presented in this work are based on reionization simulations which
do not resolve the smallest atomically-cooling halos, with masses from
$\sim10^8$ to $\sim2\times10^9\,M_\odot$. Smaller-box, higher-resolution
radiative transfer simulations which included these halos
\citet{selfregulated} showed that the presence of low-mass sources results in
self-regulation of the reionization process, whereby $\tau_{\rm es}$ is
boosted, while the large-scale structure of reionization and its end are not
affected. The last feature is due to the strong suppression of these low-mass
sources due to Jeans-mass filtering in the ionized regions. We expect that
this self-regulation would not affect our current results significantly since
the CMB polarization signal from the patchiness is dominated by the large
bubbles. We should expect some boost in the power due to the higher optical
depth and more extended nature of these reionization scenarios. The full
calculation of this effect is still difficult, however, since the small
computational box sizes currently required in order to resolve the low-mass
sources would result in a large cosmic variance, in addition to
underestimating the large-scale power of the ionization fluctuations.

\acknowledgements

We thank Wayne Hu and Michael Mortonson for constructive exchanges.

\bibliography{polsim}

\begin{thebibliography}{36}
\expandafter\ifx\csname natexlab\endcsname\relax\def\natexlab#1{#1}\fi
\expandafter\ifx\csname bibnamefont\endcsname\relax
  \def\bibnamefont#1{#1}\fi
\expandafter\ifx\csname bibfnamefont\endcsname\relax
  \def\bibfnamefont#1{#1}\fi
\expandafter\ifx\csname citenamefont\endcsname\relax
  \def\citenamefont#1{#1}\fi
\expandafter\ifx\csname url\endcsname\relax
  \def\url#1{\texttt{#1}}\fi
\expandafter\ifx\csname urlprefix\endcsname\relax\def\urlprefix{URL }\fi
\providecommand{\bibinfo}[2]{#2}
\providecommand{\eprint}[2][]{\url{#2}}

\bibitem[{\citenamefont{{Page} et~al.}(2006)\citenamefont{{Page}, {Hinshaw},
  {Komatsu}, {Nolta}, {Spergel}, {Bennett}, {Barnes}, {Bean}, {Dore'},
  {Halpern} et~al.}}]{page06}
\bibinfo{author}{\bibfnamefont{L.}~\bibnamefont{{Page}}},
  \bibinfo{author}{\bibfnamefont{G.}~\bibnamefont{{Hinshaw}}},
  \bibinfo{author}{\bibfnamefont{E.}~\bibnamefont{{Komatsu}}},
  \bibinfo{author}{\bibfnamefont{M.~R.} \bibnamefont{{Nolta}}},
  \bibinfo{author}{\bibfnamefont{D.~N.} \bibnamefont{{Spergel}}},
  \bibinfo{author}{\bibfnamefont{C.~L.} \bibnamefont{{Bennett}}},
  \bibinfo{author}{\bibfnamefont{C.}~\bibnamefont{{Barnes}}},
  \bibinfo{author}{\bibfnamefont{R.}~\bibnamefont{{Bean}}},
  \bibinfo{author}{\bibfnamefont{O.}~\bibnamefont{{Dore'}}},
  \bibinfo{author}{\bibfnamefont{M.}~\bibnamefont{{Halpern}}},
  \bibnamefont{et~al.}, \bibinfo{journal}{ArXiv Astrophysics e-prints}
  (\bibinfo{year}{2006}), \eprint{astro-ph/0603450}.

\bibitem[{\citenamefont{{Weller}}(1999)}]{weller99}
\bibinfo{author}{\bibfnamefont{J.}~\bibnamefont{{Weller}}},
  \bibinfo{journal}{\apjl} \textbf{\bibinfo{volume}{527}}, \bibinfo{pages}{L1}
  (\bibinfo{year}{1999}), \eprint{astro-ph/9908033}.

\bibitem[{\citenamefont{{Hu}}(2000)}]{hu00}
\bibinfo{author}{\bibfnamefont{W.}~\bibnamefont{{Hu}}}, \bibinfo{journal}{\apj}
  \textbf{\bibinfo{volume}{529}}, \bibinfo{pages}{12} (\bibinfo{year}{2000}),
  \eprint{astro-ph/9907103}.

\bibitem[{\citenamefont{{Liu} et~al.}(2001)\citenamefont{{Liu}, {Sugiyama},
  {Benson}, {Lacey}, and {Nusser}}}]{liu01}
\bibinfo{author}{\bibfnamefont{G.-C.} \bibnamefont{{Liu}}},
  \bibinfo{author}{\bibfnamefont{N.}~\bibnamefont{{Sugiyama}}},
  \bibinfo{author}{\bibfnamefont{A.~J.} \bibnamefont{{Benson}}},
  \bibinfo{author}{\bibfnamefont{C.~G.} \bibnamefont{{Lacey}}},
  \bibnamefont{and} \bibinfo{author}{\bibfnamefont{A.}~\bibnamefont{{Nusser}}},
  \bibinfo{journal}{\apj} \textbf{\bibinfo{volume}{561}}, \bibinfo{pages}{504}
  (\bibinfo{year}{2001}), \eprint{astro-ph/0101368}.

\bibitem[{\citenamefont{{Mortonson} and {Hu}}(2006)}]{mortonson06}
\bibinfo{author}{\bibfnamefont{M.~J.} \bibnamefont{{Mortonson}}}
  \bibnamefont{and} \bibinfo{author}{\bibfnamefont{W.}~\bibnamefont{{Hu}}},
  \bibinfo{journal}{ArXiv Astrophysics e-prints}  (\bibinfo{year}{2006}),
  \eprint{astro-ph/0607652}.

\bibitem[{\citenamefont{{Mellema}
  et~al.}(2006{\natexlab{a}})\citenamefont{{Mellema}, {Iliev}, {Pen}, and
  {Shapiro}}}]{21cmreionpaper}
\bibinfo{author}{\bibfnamefont{G.}~\bibnamefont{{Mellema}}},
  \bibinfo{author}{\bibfnamefont{I.~T.} \bibnamefont{{Iliev}}},
  \bibinfo{author}{\bibfnamefont{U.-L.} \bibnamefont{{Pen}}}, \bibnamefont{and}
  \bibinfo{author}{\bibfnamefont{P.~R.} \bibnamefont{{Shapiro}}},
  \bibinfo{journal}{\mnras} \textbf{\bibinfo{volume}{372}},
  \bibinfo{pages}{679} (\bibinfo{year}{2006}{\natexlab{a}}),
  \eprint{astro-ph/0603518}.

\bibitem[{\citenamefont{{Iliev}
  et~al.}(2006{\natexlab{a}})\citenamefont{{Iliev}, {Pen}, {Richard Bond},
  {Mellema}, and {Shapiro}}}]{2006NewAR..50..909I}
\bibinfo{author}{\bibfnamefont{I.~T.} \bibnamefont{{Iliev}}},
  \bibinfo{author}{\bibfnamefont{U.-L.} \bibnamefont{{Pen}}},
  \bibinfo{author}{\bibfnamefont{J.}~\bibnamefont{{Richard Bond}}},
  \bibinfo{author}{\bibfnamefont{G.}~\bibnamefont{{Mellema}}},
  \bibnamefont{and} \bibinfo{author}{\bibfnamefont{P.~R.}
  \bibnamefont{{Shapiro}}}, \bibinfo{journal}{New Astronomy Review}
  \textbf{\bibinfo{volume}{50}}, \bibinfo{pages}{909}
  (\bibinfo{year}{2006}{\natexlab{a}}), \eprint{astro-ph/0607209}.

\bibitem[{\citenamefont{{Iliev}
  et~al.}(2006{\natexlab{b}})\citenamefont{{Iliev}, {Pen}, {Bond}, {Mellema},
  and {Shapiro}}}]{kSZ}
\bibinfo{author}{\bibfnamefont{I.~T.} \bibnamefont{{Iliev}}},
  \bibinfo{author}{\bibfnamefont{U.-L.} \bibnamefont{{Pen}}},
  \bibinfo{author}{\bibfnamefont{J.~R.} \bibnamefont{{Bond}}},
  \bibinfo{author}{\bibfnamefont{G.}~\bibnamefont{{Mellema}}},
  \bibnamefont{and} \bibinfo{author}{\bibfnamefont{P.~R.}
  \bibnamefont{{Shapiro}}}, \bibinfo{journal}{submitted to \mnras}
  (\bibinfo{year}{2006}{\natexlab{b}}), \eprint{astro-ph/0609592}.

\bibitem[{\citenamefont{{Knox} and {Song}}(2002)}]{knox02}
\bibinfo{author}{\bibfnamefont{L.}~\bibnamefont{{Knox}}} \bibnamefont{and}
  \bibinfo{author}{\bibfnamefont{Y.-S.} \bibnamefont{{Song}}},
  \bibinfo{journal}{Physical Review Letters} \textbf{\bibinfo{volume}{89}},
  \bibinfo{pages}{011303} (\bibinfo{year}{2002}), \eprint{astro-ph/0202286}.

\bibitem[{\citenamefont{{Seljak} and {Hirata}}(2004)}]{seljak05}
\bibinfo{author}{\bibfnamefont{U.}~\bibnamefont{{Seljak}}} \bibnamefont{and}
  \bibinfo{author}{\bibfnamefont{C.~M.} \bibnamefont{{Hirata}}},
  \bibinfo{journal}{\prd} \textbf{\bibinfo{volume}{69}},
  \bibinfo{pages}{043005} (\bibinfo{year}{2004}), \eprint{astro-ph/0310163}.

\bibitem[{\citenamefont{{Iliev}
  et~al.}(2006{\natexlab{c}})\citenamefont{{Iliev}, {Mellema}, {Pen}, {Merz},
  {Shapiro}, and {Alvarez}}}]{2006MNRAS.369.1625I}
\bibinfo{author}{\bibfnamefont{I.~T.} \bibnamefont{{Iliev}}},
  \bibinfo{author}{\bibfnamefont{G.}~\bibnamefont{{Mellema}}},
  \bibinfo{author}{\bibfnamefont{U.-L.} \bibnamefont{{Pen}}},
  \bibinfo{author}{\bibfnamefont{H.}~\bibnamefont{{Merz}}},
  \bibinfo{author}{\bibfnamefont{P.~R.} \bibnamefont{{Shapiro}}},
  \bibnamefont{and} \bibinfo{author}{\bibfnamefont{M.~A.}
  \bibnamefont{{Alvarez}}}, \bibinfo{journal}{\mnras}
  \textbf{\bibinfo{volume}{369}}, \bibinfo{pages}{1625}
  (\bibinfo{year}{2006}{\natexlab{c}}), \eprint{astro-ph/0512187}.

\bibitem[{\citenamefont{{Mellema}
  et~al.}(2006{\natexlab{b}})\citenamefont{{Mellema}, {Iliev}, {Pen}, and
  {Shapiro}}}]{mellema06}
\bibinfo{author}{\bibfnamefont{G.}~\bibnamefont{{Mellema}}},
  \bibinfo{author}{\bibfnamefont{I.~T.} \bibnamefont{{Iliev}}},
  \bibinfo{author}{\bibfnamefont{U.-L.} \bibnamefont{{Pen}}}, \bibnamefont{and}
  \bibinfo{author}{\bibfnamefont{P.~R.} \bibnamefont{{Shapiro}}},
  \bibinfo{journal}{\mnras} \textbf{\bibinfo{volume}{372}},
  \bibinfo{pages}{679} (\bibinfo{year}{2006}{\natexlab{b}}),
  \eprint{astro-ph/0603518}.

\bibitem[{\citenamefont{{Iliev}
  et~al.}(2006{\natexlab{d}})\citenamefont{{Iliev}, {Mellema}, {Shapiro}, and
  {Pen}}}]{selfregulated}
\bibinfo{author}{\bibfnamefont{I.~T.} \bibnamefont{{Iliev}}},
  \bibinfo{author}{\bibfnamefont{G.}~\bibnamefont{{Mellema}}},
  \bibinfo{author}{\bibfnamefont{P.~R.} \bibnamefont{{Shapiro}}},
  \bibnamefont{and} \bibinfo{author}{\bibfnamefont{U.-L.} \bibnamefont{{Pen}}},
  \bibinfo{journal}{MNRAS in press}  (\bibinfo{year}{2006}{\natexlab{d}}),
  \eprint{astro-ph/0607517}.

\bibitem[{\citenamefont{{Merz} et~al.}(2005)\citenamefont{{Merz}, {Pen}, and
  {Trac}}}]{2005NewA...10..393M}
\bibinfo{author}{\bibfnamefont{H.}~\bibnamefont{{Merz}}},
  \bibinfo{author}{\bibfnamefont{U.-L.} \bibnamefont{{Pen}}}, \bibnamefont{and}
  \bibinfo{author}{\bibfnamefont{H.}~\bibnamefont{{Trac}}},
  \bibinfo{journal}{New Astronomy} \textbf{\bibinfo{volume}{10}},
  \bibinfo{pages}{393} (\bibinfo{year}{2005}).

\bibitem[{\citenamefont{{Spergel}~\etal}(2003)}]{2003ApJS..148..175S}
\bibinfo{author}{\bibfnamefont{D.~N.} \bibnamefont{{Spergel}~\etal}},
  \bibinfo{journal}{\apjs} \textbf{\bibinfo{volume}{148}}, \bibinfo{pages}{175}
  (\bibinfo{year}{2003}).

\bibitem[{\citenamefont{{Spergel}~\etal}(2006)}]{2006astro.ph..3449S}
\bibinfo{author}{\bibfnamefont{D.~N.} \bibnamefont{{Spergel}~\etal}},
  \bibinfo{journal}{ArXiv Astrophysics e-prints (arXiv:astro-ph/0603449)}
  (\bibinfo{year}{2006}), \eprint{arXiv:astro-ph/0603449}.

\bibitem[{\citenamefont{{Seljak} and
  {Zaldarriaga}}(1996)}]{1996ApJ...469..437S}
\bibinfo{author}{\bibfnamefont{U.}~\bibnamefont{{Seljak}}} \bibnamefont{and}
  \bibinfo{author}{\bibfnamefont{M.}~\bibnamefont{{Zaldarriaga}}},
  \bibinfo{journal}{\apj} \textbf{\bibinfo{volume}{469}}, \bibinfo{pages}{437}
  (\bibinfo{year}{1996}).

\bibitem[{\citenamefont{{Mellema}
  et~al.}(2006{\natexlab{c}})\citenamefont{{Mellema}, {Iliev}, {Alvarez}, and
  {Shapiro}}}]{methodpaper}
\bibinfo{author}{\bibfnamefont{G.}~\bibnamefont{{Mellema}}},
  \bibinfo{author}{\bibfnamefont{I.~T.} \bibnamefont{{Iliev}}},
  \bibinfo{author}{\bibfnamefont{M.~A.} \bibnamefont{{Alvarez}}},
  \bibnamefont{and} \bibinfo{author}{\bibfnamefont{P.~R.}
  \bibnamefont{{Shapiro}}}, \bibinfo{journal}{New Astronomy}
  \textbf{\bibinfo{volume}{11}}, \bibinfo{pages}{374}
  (\bibinfo{year}{2006}{\natexlab{c}}).

\bibitem[{\citenamefont{{Iliev}
  et~al.}(2006{\natexlab{e}})\citenamefont{{Iliev}, {Ciardi}, {Alvarez},
  {Maselli}, {Ferrara}, {Gnedin}, {Mellema}, {Nakamoto}, {Norman}, {Razoumov}
  et~al.}}]{2006MNRAS.371.1057I}
\bibinfo{author}{\bibfnamefont{I.~T.} \bibnamefont{{Iliev}}},
  \bibinfo{author}{\bibfnamefont{B.}~\bibnamefont{{Ciardi}}},
  \bibinfo{author}{\bibfnamefont{M.~A.} \bibnamefont{{Alvarez}}},
  \bibinfo{author}{\bibfnamefont{A.}~\bibnamefont{{Maselli}}},
  \bibinfo{author}{\bibfnamefont{A.}~\bibnamefont{{Ferrara}}},
  \bibinfo{author}{\bibfnamefont{N.~Y.} \bibnamefont{{Gnedin}}},
  \bibinfo{author}{\bibfnamefont{G.}~\bibnamefont{{Mellema}}},
  \bibinfo{author}{\bibfnamefont{T.}~\bibnamefont{{Nakamoto}}},
  \bibinfo{author}{\bibfnamefont{M.~L.} \bibnamefont{{Norman}}},
  \bibinfo{author}{\bibfnamefont{A.~O.} \bibnamefont{{Razoumov}}},
  \bibnamefont{et~al.}, \bibinfo{journal}{\mnras}
  \textbf{\bibinfo{volume}{371}}, \bibinfo{pages}{1057}
  (\bibinfo{year}{2006}{\natexlab{e}}), \eprint{astro-ph/0603199}.

\bibitem[{\citenamefont{{Shapiro} et~al.}(2004)\citenamefont{{Shapiro},
  {Iliev}, and {Raga}}}]{2004MNRAS.348..753S}
\bibinfo{author}{\bibfnamefont{P.~R.} \bibnamefont{{Shapiro}}},
  \bibinfo{author}{\bibfnamefont{I.~T.} \bibnamefont{{Iliev}}},
  \bibnamefont{and} \bibinfo{author}{\bibfnamefont{A.~C.}
  \bibnamefont{{Raga}}}, \bibinfo{journal}{\mnras}
  \textbf{\bibinfo{volume}{348}}, \bibinfo{pages}{753} (\bibinfo{year}{2004}).

\bibitem[{\citenamefont{{Iliev}
  et~al.}(2005{\natexlab{a}})\citenamefont{{Iliev}, {Shapiro}, and
  {Raga}}}]{2005MNRAS...361..405I}
\bibinfo{author}{\bibfnamefont{I.~T.} \bibnamefont{{Iliev}}},
  \bibinfo{author}{\bibfnamefont{P.~R.} \bibnamefont{{Shapiro}}},
  \bibnamefont{and} \bibinfo{author}{\bibfnamefont{A.~C.}
  \bibnamefont{{Raga}}}, \bibinfo{journal}{\mnras}
  \textbf{\bibinfo{volume}{361}}, \bibinfo{pages}{405}
  (\bibinfo{year}{2005}{\natexlab{a}}).

\bibitem[{\citenamefont{{Iliev}
  et~al.}(2005{\natexlab{b}})\citenamefont{{Iliev}, {Scannapieco}, and
  {Shapiro}}}]{2005ApJ...624..491I}
\bibinfo{author}{\bibfnamefont{I.~T.} \bibnamefont{{Iliev}}},
  \bibinfo{author}{\bibfnamefont{E.}~\bibnamefont{{Scannapieco}}},
  \bibnamefont{and} \bibinfo{author}{\bibfnamefont{P.~R.}
  \bibnamefont{{Shapiro}}}, \bibinfo{journal}{\apj}
  \textbf{\bibinfo{volume}{624}}, \bibinfo{pages}{491}
  (\bibinfo{year}{2005}{\natexlab{b}}).

\bibitem[{\citenamefont{{Ciardi} et~al.}(2006)\citenamefont{{Ciardi},
  {Scannapieco}, {Stoehr}, {Ferrara}, {Iliev}, and {Shapiro}}}]{MH_sim}
\bibinfo{author}{\bibfnamefont{B.}~\bibnamefont{{Ciardi}}},
  \bibinfo{author}{\bibfnamefont{E.}~\bibnamefont{{Scannapieco}}},
  \bibinfo{author}{\bibfnamefont{F.}~\bibnamefont{{Stoehr}}},
  \bibinfo{author}{\bibfnamefont{A.}~\bibnamefont{{Ferrara}}},
  \bibinfo{author}{\bibfnamefont{I.~T.} \bibnamefont{{Iliev}}},
  \bibnamefont{and} \bibinfo{author}{\bibfnamefont{P.~R.}
  \bibnamefont{{Shapiro}}}, \bibinfo{journal}{MNRAS}
  \textbf{\bibinfo{volume}{366}}, \bibinfo{pages}{689} (\bibinfo{year}{2006}).

\bibitem[{\citenamefont{{McQuinn} et~al.}(2006)\citenamefont{{McQuinn}, {Lidz},
  {Zahn}, {Dutta}, {Hernquist}, and {Zaldarriaga}}}]{2006astro.ph.10094M}
\bibinfo{author}{\bibfnamefont{M.}~\bibnamefont{{McQuinn}}},
  \bibinfo{author}{\bibfnamefont{A.}~\bibnamefont{{Lidz}}},
  \bibinfo{author}{\bibfnamefont{O.}~\bibnamefont{{Zahn}}},
  \bibinfo{author}{\bibfnamefont{S.}~\bibnamefont{{Dutta}}},
  \bibinfo{author}{\bibfnamefont{L.}~\bibnamefont{{Hernquist}}},
  \bibnamefont{and}
  \bibinfo{author}{\bibfnamefont{M.}~\bibnamefont{{Zaldarriaga}}},
  \bibinfo{journal}{ArXiv Astrophysics e-prints}  (\bibinfo{year}{2006}),
  \eprint{astro-ph/0610094}.

\bibitem[{\citenamefont{{Zaldarriaga} and {Seljak}}(1997)}]{zaldarriaga97}
\bibinfo{author}{\bibfnamefont{M.}~\bibnamefont{{Zaldarriaga}}}
  \bibnamefont{and} \bibinfo{author}{\bibfnamefont{U.}~\bibnamefont{{Seljak}}},
  \bibinfo{journal}{\prd} \textbf{\bibinfo{volume}{55}}, \bibinfo{pages}{1830}
  (\bibinfo{year}{1997}), \eprint{astro-ph/9609170}.

\bibitem[{\citenamefont{{Amblard} and {White}}(2005)}]{amblard05}
\bibinfo{author}{\bibfnamefont{A.}~\bibnamefont{{Amblard}}} \bibnamefont{and}
  \bibinfo{author}{\bibfnamefont{M.}~\bibnamefont{{White}}},
  \bibinfo{journal}{New Astronomy} \textbf{\bibinfo{volume}{10}},
  \bibinfo{pages}{417} (\bibinfo{year}{2005}), \eprint{astro-ph/0409063}.

\bibitem[{\citenamefont{{Portsmouth}}(2004)}]{portsmouth04}
\bibinfo{author}{\bibfnamefont{J.}~\bibnamefont{{Portsmouth}}},
  \bibinfo{journal}{\prd} \textbf{\bibinfo{volume}{70}},
  \bibinfo{pages}{063504} (\bibinfo{year}{2004}), \eprint{astro-ph/0402173}.

\bibitem[{\citenamefont{{Hu} and {White}}(1997)}]{hu97}
\bibinfo{author}{\bibfnamefont{W.}~\bibnamefont{{Hu}}} \bibnamefont{and}
  \bibinfo{author}{\bibfnamefont{M.}~\bibnamefont{{White}}},
  \bibinfo{journal}{\prd} \textbf{\bibinfo{volume}{56}}, \bibinfo{pages}{596}
  (\bibinfo{year}{1997}), \eprint{astro-ph/9702170}.

\bibitem[{\citenamefont{{Okamoto} and {Hu}}(2002)}]{okamoto02}
\bibinfo{author}{\bibfnamefont{T.}~\bibnamefont{{Okamoto}}} \bibnamefont{and}
  \bibinfo{author}{\bibfnamefont{W.}~\bibnamefont{{Hu}}},
  \bibinfo{journal}{\prd} \textbf{\bibinfo{volume}{66}},
  \bibinfo{pages}{063008} (\bibinfo{year}{2002}), \eprint{astro-ph/0206155}.

\bibitem[{\citenamefont{{Baumann} and {Cooray}}(2003)}]{baumann03b}
\bibinfo{author}{\bibfnamefont{D.}~\bibnamefont{{Baumann}}} \bibnamefont{and}
  \bibinfo{author}{\bibfnamefont{A.}~\bibnamefont{{Cooray}}},
  \bibinfo{journal}{New Astronomy Review} \textbf{\bibinfo{volume}{47}},
  \bibinfo{pages}{839} (\bibinfo{year}{2003}), \eprint{astro-ph/0304416}.

\bibitem[{\citenamefont{{Yadav} and {Wandelt}}(2005)}]{yadav05}
\bibinfo{author}{\bibfnamefont{A.~P.} \bibnamefont{{Yadav}}} \bibnamefont{and}
  \bibinfo{author}{\bibfnamefont{B.~D.} \bibnamefont{{Wandelt}}},
  \bibinfo{journal}{\prd} \textbf{\bibinfo{volume}{71}},
  \bibinfo{pages}{123004} (\bibinfo{year}{2005}), \eprint{astro-ph/0505386}.

\bibitem[{\citenamefont{{Holder} et~al.}(2006)\citenamefont{{Holder}, {Iliev},
  and {Mellema}}}]{pol21}
\bibinfo{author}{\bibfnamefont{G.}~\bibnamefont{{Holder}}},
  \bibinfo{author}{\bibfnamefont{I.~T.} \bibnamefont{{Iliev}}},
  \bibnamefont{and}
  \bibinfo{author}{\bibfnamefont{G.}~\bibnamefont{{Mellema}}},
  \bibinfo{journal}{ArXiv Astrophysics e-prints}  (\bibinfo{year}{2006}),
  \eprint{astro-ph/0609689}.

\bibitem[{\citenamefont{{Kaiser}}(1992)}]{kaiser92}
\bibinfo{author}{\bibfnamefont{N.}~\bibnamefont{{Kaiser}}},
  \bibinfo{journal}{\apj} \textbf{\bibinfo{volume}{388}}, \bibinfo{pages}{272}
  (\bibinfo{year}{1992}).

\bibitem[{\citenamefont{{Limber}}(1953)}]{limber53}
\bibinfo{author}{\bibfnamefont{D.~N.} \bibnamefont{{Limber}}},
  \bibinfo{journal}{\apj} \textbf{\bibinfo{volume}{117}}, \bibinfo{pages}{134}
  (\bibinfo{year}{1953}).

\bibitem[{\citenamefont{{Kosowsky}}(2006)}]{kosowsky06}
\bibinfo{author}{\bibfnamefont{A.}~\bibnamefont{{Kosowsky}}},
  \bibinfo{journal}{New Astronomy Review} \textbf{\bibinfo{volume}{50}},
  \bibinfo{pages}{969} (\bibinfo{year}{2006}), \eprint{astro-ph/0608549}.

\bibitem[{\citenamefont{{Ruhl} et~al.}(2004)\citenamefont{{Ruhl}, {Ade},
  {Carlstrom}, {Cho}, {Crawford}, {Dobbs}, {Greer}, {Halverson}, {Holzapfel},
  {Lanting} et~al.}}]{ruhl04}
\bibinfo{author}{\bibfnamefont{J.}~\bibnamefont{{Ruhl}}},
  \bibinfo{author}{\bibfnamefont{P.~A.~R.} \bibnamefont{{Ade}}},
  \bibinfo{author}{\bibfnamefont{J.~E.} \bibnamefont{{Carlstrom}}},
  \bibinfo{author}{\bibfnamefont{H.-M.} \bibnamefont{{Cho}}},
  \bibinfo{author}{\bibfnamefont{T.}~\bibnamefont{{Crawford}}},
  \bibinfo{author}{\bibfnamefont{M.}~\bibnamefont{{Dobbs}}},
  \bibinfo{author}{\bibfnamefont{C.~H.} \bibnamefont{{Greer}}},
  \bibinfo{author}{\bibfnamefont{N.~w.} \bibnamefont{{Halverson}}},
  \bibinfo{author}{\bibfnamefont{W.~L.} \bibnamefont{{Holzapfel}}},
  \bibinfo{author}{\bibfnamefont{T.~M.} \bibnamefont{{Lanting}}},
  \bibnamefont{et~al.}, in \emph{\bibinfo{booktitle}{Z-Spec: a broadband
  millimeter-wave grating spectrometer: design, construction, and first
  cryogenic measurements. Edited by Bradford, C. Matt; Ade, Peter A. R.;
  Aguirre, James E.; Bock, James J.; Dragovan, Mark; Duband, Lionel; Earle,
  Lieko; Glenn, Jason; Matsuhara, Hideo; Naylor, Bret J.; Nguyen, Hien T.; Yun,
  Minhee; Zmuidzinas, Jonas. Proceedings of the SPIE, Volume 5498, pp. 11-29
  (2004).}}, edited by \bibinfo{editor}{\bibfnamefont{C.~M.}
  \bibnamefont{{Bradford}}}, \bibinfo{editor}{\bibfnamefont{P.~A.~R.}
  \bibnamefont{{Ade}}}, \bibinfo{editor}{\bibfnamefont{J.~E.}
  \bibnamefont{{Aguirre}}}, \bibinfo{editor}{\bibfnamefont{J.~J.}
  \bibnamefont{{Bock}}},
  \bibinfo{editor}{\bibfnamefont{M.}~\bibnamefont{{Dragovan}}},
  \bibinfo{editor}{\bibfnamefont{L.}~\bibnamefont{{Duband}}},
  \bibinfo{editor}{\bibfnamefont{L.}~\bibnamefont{{Earle}}},
  \bibinfo{editor}{\bibfnamefont{J.}~\bibnamefont{{Glenn}}},
  \bibinfo{editor}{\bibfnamefont{H.}~\bibnamefont{{Matsuhara}}},
  \bibinfo{editor}{\bibfnamefont{B.~J.} \bibnamefont{{Naylor}}},
  \bibnamefont{et~al.} (\bibinfo{year}{2004}), pp. \bibinfo{pages}{11--29}.

\end{thebibliography}

\appendix

\end{document}